\documentclass[universe,article,submit,pdftex,moreauthors]{Definitions/mdpi} 
\firstpage{1} 
\pubvolume{1}
\issuenum{1}
\articlenumber{0}
\pubyear{2023}
\copyrightyear{2023}
\datereceived{ } 
\daterevised{ } 
\dateaccepted{ } 
\datepublished{ } 
\hreflink{https://doi.org/} 

\Title{The apparent tidal decay of WASP-4\,b can be explained by the R\o{}mer effect}

\TitleCitation{The apparent tidal decay of WASP-4\,b can be explained by the R\o{}mer effect}

\Author{Jan-Vincent Harre $^{1}$\orcidA{} and Alexis M. S. Smith $^{1}$\orcidB{}}

\AuthorNames{Jan-Vincent Harre and Alexis M. S. Smith}

\AuthorCitation{Harre, J.-V.; Smith, A. M. S.}

\address{%
$^{1}$ \quad Institute of Planetary Research, German Aerospace Center (DLR), Rutherfordstraße 2, 12489 Berlin, Germany}

\corres{Correspondence: jan-vincent.harre@dlr.de}

\abstract{
    Tidal orbital decay plays a vital role in the evolution of hot Jupiter systems. As of now, this was only observationally confirmed for the WASP-12 system. There are a few other candidates, including WASP-4 b, but no conclusive result could be obtained for these systems as of yet.
    In this study, we present an analysis of new TESS data of WASP-4 b together with archival data, taking the light-time effect (LTE), induced by the second planetary companion, into account as well.
    We make use of three different Markov-Chain-Monte-Carlo models; a circular orbit with a constant orbital period, a circular orbit with a decaying orbit, and an elliptical orbit with apsidal precession. This analysis is repeated for four cases. The first case features no LTE correction, with the remaining three cases featuring three different timing correction approaches.
    Comparison of these models yields no conclusive answer to the cause of WASP-4\,b's apparent transit timing variations. A broad range of values of the orbital decay and apsidal precession parameters are possible, depending on the LTE correction. This work highlights the importance of continued photometric and spectroscopic monitoring of hot Jupiters.    
}


\begin{document}


\section{Introduction}
The study of exoplanets has unveiled a great diversity in terms of their physical properties, orbital characteristics, and formation mechanisms. Among the  most intriguing exoplanet classes are the hot Jupiters. These relatively rare companions (e.g. \citep{2022MNRAS.516...75B}) stand out as a group of gas giants that orbit their host stars at exceptionally close distances and have challenged our understanding of planetary formation and evolution. One of the key phenomena that has piqued the interest of astronomers is the tidal orbital decay of Hot Jupiters, where as of now only WASP-12\,b could be observationally confirmed to experience this effect \citep{2020ApJ...888L...5Y, 2021AJ....161...72T, 2022AJ....163..175W}. However, there seems to be evidence for the occurrence rates of hot Jupiters in orbit around Sun-like stars to decrease with stellar age, as opposed to the occurrence rates of cold Jupiters, as found by \citet{2023arXiv230914605M}. This is supported by the likely observation of planetary engulfment by a Sun-like star made by the Zwicky Transient Facility \citep{2023Natur.617...55D}.

Due to the close proximity of hot Jupiters to their stellar hosts, there are strong tidal interactions between the two bodies. 
These gravitational interactions can manifest themselves in the planet raising a bulge on the surface of the star. Depending on the rotational period of the star and the planetary orbital period, the viscosity of the stellar plasma can lead to a lag between the position of the tidal bulge and the virtual line connecting the stellar and planetary centers. If the star is rotating slower than the planet is orbiting around it, orbital angular momentum of the planet will be transferred to the star (equilibrium tide). The dynamical tide, arising from stellar oscillations, is also contributing to this \citep{2014ARA&A..52..171O}. This means that the star will spin up and the planetary orbit shrink gradually \citep{1973ApJ...180..307C, 1996ApJ...470.1187R}. This provides us with insights into the long-term stability of these system.

WASP-4\,b is a hot Jupiter discovered by \citet{2008ApJ...675L.113W}. It shows TTVs that have been examined before, and other effects, mimicking the orbital decay signature, were ruled out previously, like the Applegate mechanism \citep{2010MNRAS.405.2037W, 2019MNRAS.490.4230S}. 
However, apsidal precession due to a small eccentricity of the planetary orbit, or TTVs arising from the R\o{}mer effect (or light-time effect, LTE hereafter, \citealt{1952ApJ...116..211I}), due to a companion in the system, have not yet been ruled out as the cause of the TTVs, even after the discovery of the companion candidate. Previous studies examining the decay rate in the case of tidal orbital decay, found values in the range from $-2.4$\,ms\,yr$^{-1}$ to $-12.6$\,ms\,yr$^{-1}$ \citep{2019AJ....157..217B,2019MNRAS.490.1294B, 2019MNRAS.490.4230S, 2020MNRAS.496L..11B, 2022AcA....72....1M,2022AJ....163..281T}, with the most recent estimate being $\dot{P} = (-6.2\pm1.2)$\,ms\,yr$^{-1}$ \citep{2023A&A...669A.124H}.

In this work, we re-examine the TTVs in this system by making use of recently acquired TESS photometry and combine them with archival data from previous works. In particular, we account for the time shift due to the LTE induced by the additional planet candidate, as discovered by \citet{2022AJ....163..281T}.
The observations are described in Sect.~\ref{sec:obs}, with our modelling and results thereof being described in Sect.~\ref{sec:methods} and \ref{sec:results}, respectively. The latter are discussed in Sect.~\ref{sec:disc} and our final conclusions can be found in Sect.~\ref{sec:conclusion}.


\section{Observations\label{sec:obs}}

For our analysis, we made use of the previously described data set in \citet{2023A&A...669A.124H} for WASP-4\,b. In short, this data set is mainly based on the homogeneous re-analysis of previously published light curves from \citet{2020MNRAS.496L..11B}, with the addition of TESS Sectors 28 and 29, as well as a re-analysis of the TESS Sector 2 data. Furthermore, in \citep{2023A&A...669A.124H}, we re-fitted the publicly available light curves from the ExoClock project \citep{2022ApJS..258...40K} and from WASP \citep{2008ApJ...675L.113W}, and added eight transit observations taken with the CHEOPS space telescope \citep{2021ExA....51..109B}. Included in this data set are also four occultation timings from the literature \citep{2011A&A...530A...5C, 2011ApJ...727...23B, 2015MNRAS.454.3002Z}. All these timings, including three different corrections, can be found in Appendix~\ref{sec:appendixB}, Table~\ref{tab:timings} for the transit timings and Table~\ref{tab:timings_occ} for the occultation timings.

To this data set, we add new TESS observations from Sector 69 at a cadence of 120\,s. We make use of the Presearch Data Conditioning Simple Aperture Photometry (PDCSAP) flux for the analysis of the light curves, which consists of data produced by the TESS Science Processing Operations Center (SPOC) at NASA Ames Research Center \citep{2016SPIE.9913E..3EJ}. This data is publicly available at MAST\footnote{\href{https://mast.stsci.edu/portal/Mashup/Clients/Mast/Portal.html}{https://mast.stsci.edu/}}.


\section{Modelling\label{sec:methods}}

\subsection{Light curve modelling}
For the analysis of the TESS transits, we made use of the Transit and Light Curve Modeller (\textit{TLCM}, \citealt{2020MNRAS.496.4442C}) as described in \citet{2023A&A...669A.124H}. In a first run, we fitted all transits together to get the combined shape of all transits to reduce the impact of stellar activity on the transit timings. The respective priors are shown in Table~\ref{tab:TLCM_params}. During a second run, we fixed the shapes of the transits to those of the combined model and fitted all transits individually, with only the transit ephemerides being free. In both cases, we used the median solution for the final results.

\subsection{Transit timing variation analysis}
For the analysis of the mid-transit times, obtained from our light curve modelling, we employ the same three models as \citet{2023A&A...669A.124H}. The first is a model assuming a circular Keplerian orbit, describing a linear ephemeris with a constant orbital period:
\begin{eqnarray}\label{eq:linear}
    t_\mathrm{tra}(N) &=& T_0 + N \, P, \\
    t_\mathrm{occ}(N) &=& T_0 + \frac{P}{2} + N \, P,
\end{eqnarray}
where $t_\mathrm{tra}(N)$ and $t_\mathrm{occ}(N)$ are the calculated mid-transit and mid-occultation timings at the epoch $N$, $T_0$ denotes the reference mid-transit time, and $P$ the planetary orbital period. This gives us two free parameters in the fit.

The second model describes the case of a decaying orbit due to the transfer of angular momentum (see e.g. \citealt{1973ApJ...180..307C, 1996ApJ...470.1187R}). These are quadratic functions with a constant change in the orbital period of the planet:
\begin{eqnarray}
    t_\mathrm{tra}(N) &=& T_0 + N \, P + \frac{1}{2}\,\frac{dP}{dN}\,N^2, \\
    t_\mathrm{occ}(N) &=& T_0 + \frac{P}{2} + N \, P + \frac{1}{2}\,\frac{dP}{dN}\,N^2.
\end{eqnarray}
This constant change in the orbital period is denoted by the decay rate $\frac{dP}{dN}$, which can be converted to the period derivative $\dot{P} = \frac{dP}{dt} = \frac{1}{P}\,\frac{dP}{dN}$. There are three free parameters ($T_0$, $P$, and $\frac{dP}{dN}$) when fitting this model. 
The period derivative is linked to the stellar tidal modified quality factor $Q_\star'$ via the constant-phase lag model, as defined in \citet{1966Icar....5..375G}:
\begin{equation}\label{eq:Pdot_Q_relation}
    \dot P = f\frac{\pi}{Q'_\star}\,\frac{M_p}{M_\star}\,\left(\frac{R_\star}{a}\right)^5\,,
\end{equation}
where $f$ denotes the tidal factor, $M_p$ and $M_\star$ the planetary and stellar masses, $R_\star$ the stellar radius, and $a$ the semi-major axis of the planetary orbit. Depending on the ratio of the planetary orbital period to the stellar rotational period and the orbital (mis-)alignment, $f$ takes different values. If the planetary orbital period is shorter than the rotation period of the star, as is the case for WASP-4\,b \citep{2013MNRAS.434...46H}, we get $f=-\frac{27}{2}$.
Refer to Sect.~4 of \citet{2023A&A...669A.124H} for a more detailed description, also including the case of inclined orbits and different planet-to-star period ratios.

The third model we are using is an apsidal precession orbit, where a small eccentricity $e$ leads to the planetary orbit to precess around the star. This can, on relatively short timescales, induce the same TTV signature as orbital decay, which makes it hard to differentiate the two models if the orbital eccentricity is only loosely constrained. This sinusoidal model follows the descriptions of \citet{1995Ap&SS.226...99G}:
\begin{eqnarray}
    t_\mathrm{tra}(N) &=& t_0 + N \, P_s - \frac{e\,P_a}{\pi}\cos\omega(N), \\
    t_\mathrm{occ}(N) &=& t_0 + \frac{P_a}{2} + N \, P_s + \frac{e\,P_a}{\pi}\cos\omega(N),
\end{eqnarray}
where $P_s$ is the sidereal period, $P_a$ the anomalistic period, and $\omega$ the argument of pericenter. The sidereal and anomalistic periods are related via:
\begin{equation}
    P_s = P_a \left( 1-\frac{1}{2\pi}\frac{d\omega}{dN} \right),
\end{equation}
with a constant change in the argument of pericenter $\frac{d\omega}{dN}$.
The relationship between $\omega$ and $N$ can be described as follows:
\begin{equation}
    \omega(N) = \omega_0 + \frac{d\omega}{dN}\,N,
\end{equation}
with $w_0$ being the argument of pericenter at the reference time $T_0$. 
In total, this model gives us five free parameters in the fitting process.

These three models are then fitted to the data via MCMC optimization using the \textit{emcee} \textit{Python} package \citep{2013PASP..125..306F}. Per model, we use 100 walkers and use a burn-in period of 10.000 steps with a total chain length of 75.000 steps.
Convergence is ensured by checking that the chains are longer than 50 times the autocorrelation time of the parameters.
The priors for our models can be found in Table~\ref{tab:MCMC}.

\subsection{Light-time effect}

\begin{figure}
    \centering
    \includegraphics[width=.6\linewidth]{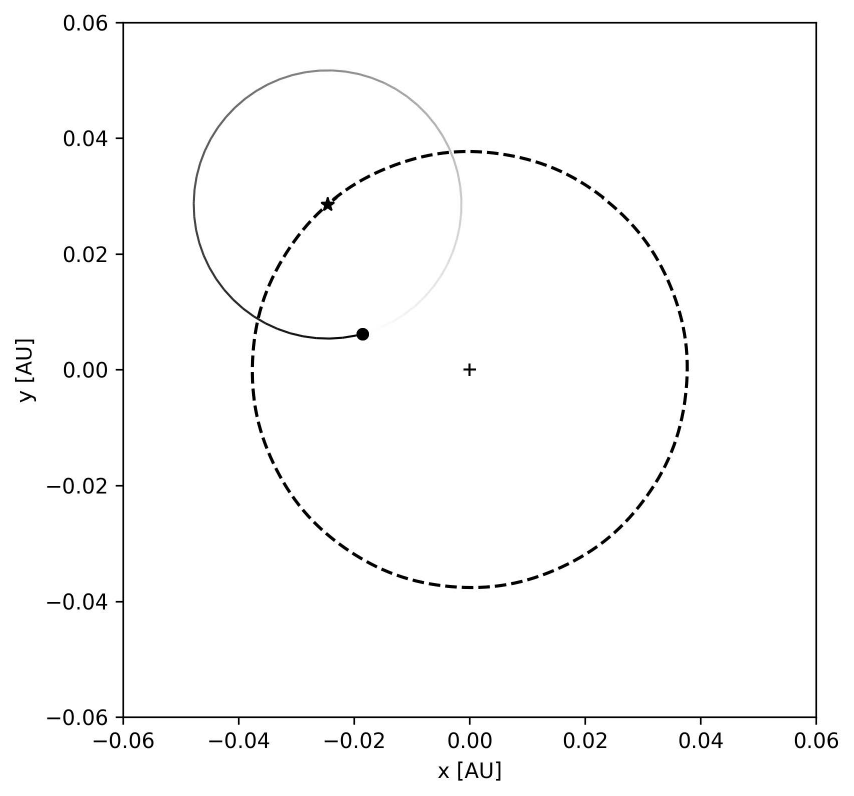}
    \caption{The orbits of WASP-4 (star symbol, dashed line) and WASP-4\,b (black dot, solid line) with reference to the system's center of mass (black plus symbol) from \textit{REBOUND}. The x- and y-axes lie within the orbital plane of the planets, which are assumed to be coplanar. The orbit of planet c is not visible in this view, since it's semi-major axis is assumed to be 6.82\,AU.}
    \label{fig:orbit_W4}
\end{figure}

The presence of a candidate planetary companion in a 7000\,d orbit, as established by \citet{2022AJ....163..281T}, would induce an orbital motion onto WASP-4. Due to the high mass of this candidate (``planet c'' hereafter), the system's center of mass is shifted by about 8.9 times the radius of WASP-4, see Fig.~\ref{fig:orbit_W4}. This will have a significant impact on the observed mid-transit times because of the time difference that it takes the light to travel from the far side to the near side of the orbit from our point of view. In some cases, this could induce TTVs akin to the imprint of tidal decay on short time scales, depending on the orbital period of planet c. To include a correction of the LTE, we apply the formula of \citet{2005ASPC..335..191S} to find the maximum time-shift from this effect for circular orbits:
\begin{equation}
    \Delta T_\mathrm{max} = 2 \frac{M_\mathrm{p}}{M_\star}\,\frac{a\sin i}{c},
\end{equation}
where $\Delta T_\mathrm{max}$ is the maximum resulting time-shift due to the LTE, $i$ is the inclination of the planetary orbit, and $c$ is the speed of light.
For WASP-4, using the planetary parameters of planet c, and assuming a circular orbit with $i=90^\circ$, we obtain a maximum time-shift of $\Delta T_\mathrm{max} = 37.7\,$s. To validate this result, we simulate the system using the N-body code \textit{REBOUND} \citep{2012A&A...537A.128R}, and measure the orbit of the star around the system's center of mass, as shown in Fig.~\ref{fig:orbit_W4}. From this, we obtain $\Delta T_\mathrm{max} = 37.6\,$s. Using the non-circular solution from \citet{2022AJ....163..281T} for planet c, we get $\Delta T_\mathrm{max,\,ell} = 41.1\,$s. However, since the non-circular solution is not preferred in their paper, we adopt the circular solution and correct our transit times according to:
\begin{equation}\label{eq:tau_t}
    \Delta T(t) =  \Delta T_\mathrm{max}\left(\frac{1}{2}\cos\left(2\pi\,\frac{t-T_{0,\mathrm{c}}}{P_\mathrm{c}}\right)+\frac{1}{2}\right),
\end{equation}
where $\Delta T(t)$ is the time-shift due to the LTE at the time $t$, $T_{0,\mathrm{c}}$ is the time of inferior conjunction of planet c, and $P_\mathrm{c} = 7001.0\,$d is the orbital period of planet c. This formula arises because the star is at its furthest point from us at the time of inferior conjunction of planet c.
Due to the uncertainty of the time of conjunction of planet c ($T_{0,\mathrm{c}} = 2455059^{+2300}_{-2100}$\,d [BJD$_\mathrm{TDB}$]), we examine three cases of the LTE correction. Firstly, assuming the best-fit value of the time of conjunction, secondly, the minimum value of the $1\sigma$ interval, and lastly, the maximum value of the $1\sigma$ interval.

\section{Results}\label{sec:results}

\subsection{Transit fitting with TLCM}
The priors for and parameters resulting from the combined TLCM fit of all TESS transits from Sector 69 to constrain the transit shape, are listed in Table~\ref{tab:TLCM_params}, with the fit of the median model to the data shown in Fig.~\ref{fig:TESS_lc}. The resulting mid-transit times can be found in Appendix~\ref{sec:appendixB} in Table~\ref{tab:timings}.

\begin{table}[]
    \centering
    \setlength{\tabcolsep}{16pt}
    \caption{The resulting parameters from the TLCM fit to the TESS S69 data. The impact parameters is given by $b$ and $u_\mathrm{a}$ and $u_\mathrm{b}$ describe the quadratic limb-darkening parameters. $T_0$ is given in [BJD$_\mathrm{TDB} - 2450000$] and $\mathcal{U}$ denotes a uniform prior.}
    \begin{tabular}{c c c}\hline
        Parameter [unit] & Prior & Result \\\hline
        $a\,R_\star^{-1}$ & $\mathcal{U}(5.4773\pm0.1000)$ & $5.39795\pm0.01864$ \\
        $R_\mathrm{P}\,R_\star^{-1}$ & $\mathcal{U}(0.1540\pm0.0100)$ & $0.1524\pm0.0009$ \\
        $b$ & $\mathcal{U}(0.5\pm1.0)$ & $0.219\pm0.015$ \\
        $P$ [d] & $\mathcal{U}(1.338231\pm0.000100)$ & $1.338239\pm0.000014$ \\
        $T_0$ & $\mathcal{U}(10192.575\pm0.050)$ & $10192.57567\pm0.00008$ \\
        $u_\mathrm{a}$ & $\mathcal{U}(0.5\pm1.0)$ & $0.35 \pm 0.10$ \\
        $u_\mathrm{b}$ & $\mathcal{U}(0.5\pm1.0)$ & $0.22 \pm 0.14$ \\\hline
    \end{tabular}
    \label{tab:TLCM_params}
\end{table}

\begin{figure}[]
    \centering
    \includegraphics[width=0.6\linewidth]{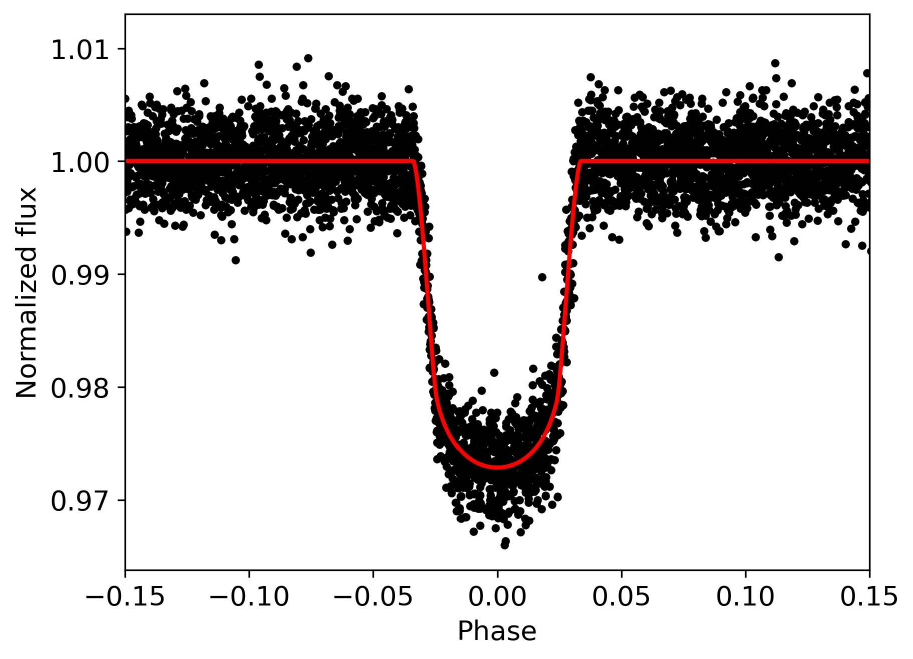}
    \caption{Phase-folded TESS light curve of WASP-4\,b from Sector 69. The data are shown as black dots after modelling and correction with \textit{TLCM}, with the median solution model shown as the red line.}
    \label{fig:TESS_lc}
\end{figure}

\subsection{TTV fits without LTE correction}
The priors for the MCMC modelling of the circular orbit, orbital decay and apsidal precession models and their resulting parameters are listed in Table~\ref{tab:MCMC}.

\begin{table*}[]
    \centering
    \setlength{\tabcolsep}{16pt}    \caption{Priors and results from our MCMC modelling of the transit timings without any LTE correction. $T_0$ is given in BJD$_\mathrm{TDB} - 2450000$. Units are given in square brackets if applicable. Priors were chosen according to the results of \citet{2023A&A...669A.124H} with enough flexibility for the walkers to sufficiently explore the parameter space.}
    \begin{adjustwidth}{-\extralength}{0cm}
    \begin{tabular}{c c c c}\hline
        Model & Parameter & Prior & Result\\\hline
        Circular orbit      & $T_0$                     & $\mathcal{U}(7000,8000)$              & $7490.68717\pm0.00002$ \\
                            & $P$ [d]                   & $\mathcal{U}(1.3,1.4)$                & $1.33823133\pm0.00000001$ \\\vspace{0.3cm}
                            & BIC                       & -                                     & 382.45 \\
        Orbital decay       & $T_0$                     & $\mathcal{U}(7000,8000)$              & $7490.68735\pm0.00002$ \\
                            & $P$ [d]                   & $\mathcal{U}(1.3,1.4)$                & $1.33823122\pm0.00000001$ \\
                            & $dP/dN$ [d/orbit]         & $\mathcal{U}(-10^{-8}, -10^{-11})$    & $(-2.43\pm0.22)\times10^{-10}$ \\\vspace{0.3cm}
                            & BIC                       & -                                     & 263.01 \\
        Apsidal precession  & $T_0$                     & $\mathcal{U}(7000,8000)$              & $7490.68690\pm0.00019$ \\
                            & $P$ [d]                   & $\mathcal{U}(1.3,1.4)$                & $1.33823140\pm0.00000009$ \\
                            & $d\omega/dN$ [rad/orbit]  & $\mathcal{U}(0.0001,0.002)$           & $(7.68\pm1.60)\times10^{-4}$ \\
                            & e                         & $\mathcal{U}(10^{-6}, 10^{-2})$       & $0.0013\pm0.0005$ \\
                            & $\omega_0$ [rad]          & $\mathcal{U}(2, 2\pi)$                   & $3.770\pm0.242$ \\
                            & BIC                       & -                                     & 271.62 \\
         \hline
    \end{tabular}
    \end{adjustwidth}
    \label{tab:MCMC}
\end{table*}

We find an orbital decay rate of $\dot{P} = \frac{dP}{dt} = (-5.75\pm0.52)$\,ms\,yr$^{-1}$, leading to a stellar modified quality factor for WASP-4 of $Q_\star' = (6.10\pm0.55)\times10^4$ and a decay timescale of $\tau = (20.2\pm1.8)$\,Myr from the median solution.
The BIC values of our solutions are obtained via:
\begin{equation}
    BIC = \chi^2 + k\cdot \ln(N),
\end{equation}
with $k$ the number of free parameters and $N$ the number of data points.
For the orbital decay fit, we find $\Delta \mathrm{BIC}_\mathrm{decay} = 119.4$ between the linear and tidal decay models, in favour of the latter.
In the case of apsidal precession, we find a precession rate of $\dot{\omega} = \frac{dw}{dt} = \frac{1}{P}\,\frac{dw}{dN} = (0.033\pm0.007)^\circ\,d^{-1}$ and a small eccentricity of $e = 0.0013\pm0.0005$. The fit yields $\Delta \mathrm{BIC}_\mathrm{precession} = 110.7$ between the linear and apsidal precession models. This leads to a difference of $\Delta \mathrm{BIC}_\mathrm{dec., prec.} = 8.7$ in favour of the orbital decay model.

\begin{figure*}
    \centering
    \includegraphics[width=\textwidth]{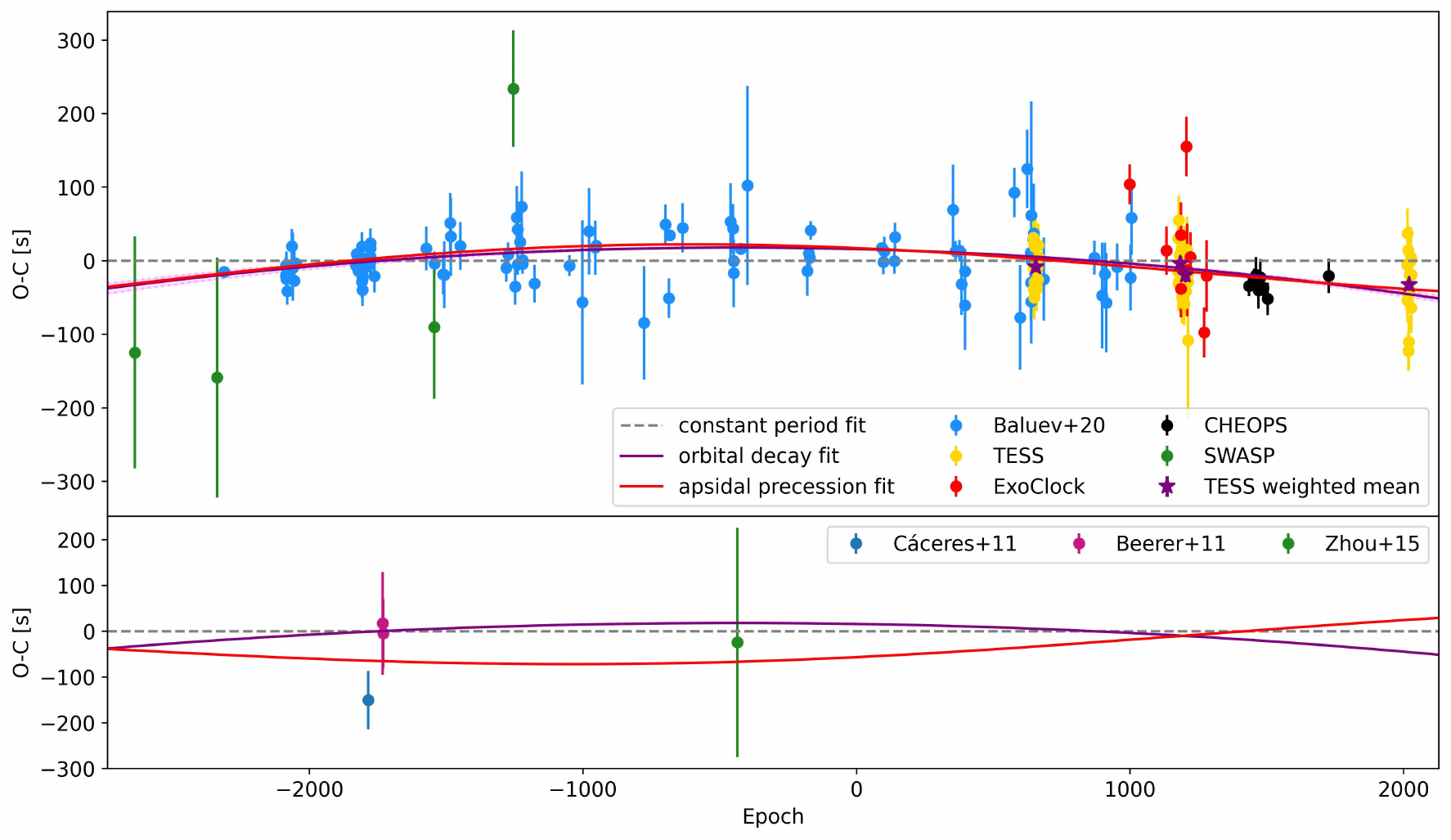}
    \caption{O-C plot showing all transit timing (top) and occultation data (bottom) together with the orbital decay and apsidal precession models. The transit number from the reference epoch is shown on the x-axis, with the deviation from the median linear ephemeris is shown on the y-axis. Colors according to the legend, with the pink shaded area showing the $1\sigma$ interval around the median solution orbital decay fit. For the available TESS data, the weighted mean timings with their respective error bars for each sector have been added.}
    \label{fig:TTVs}
\end{figure*}

The result of the MCMC fit to the transit and occultation timing data including the final median models is shown in Fig.~\ref{fig:TTVs}. Especially the latest data points highlight the deviation from a linear ephemeris in this system.

\subsection{TTV fits with LTE correction}

In case the candidate planet c does exist, we corrected for the induced LTE with a maximum time-shift of about 38\,s in three different cases. These cases are (1) planet c has the best-fit time of inferior conjunction from model \#3 of \citet{2022AJ....163..281T}, (2) the time of inferior conjunction is at the lower boundary of the respective 1$\sigma$ confidence interval, and (3) the time of inferior conjunction is at the upper boundary of the 1$\sigma$ confidence interval. The corrections, according to Eq.~\ref{eq:tau_t}, are subtracted for each transit and occultation timing.

\begin{table*}[]
    \centering
    \setlength{\tabcolsep}{7pt}
    \caption{Results from our MCMC modelling for the three cases of the LTE correction. The priors are the same as those given in Table~\ref{tab:MCMC}. $T_0$ is given in BJD$_\mathrm{TDB} - 2450000$. Units are given in square brackets if applicable. LTE refers to case (1), LTE$_\mathrm{low}$ and LTE$_\mathrm{up}$ refer to cases (2) and (3), respectively. $\Delta$BIC is defined relative to the circular Keplerian orbit case for each LTE correction. Note that the $\Delta$BIC values should only be compared within a column, not between columns.}
    \begin{adjustwidth}{-\extralength}{0cm}
    \begin{tabular}{c c c c c}\hline
        Model & Parameter                       & Result LTE                        & Result LTE$_\mathrm{lower}$       & Result LTE$_\mathrm{upper}$\\\hline
        Circ. Orbit  & $T_0$                    &  $7490.68699\pm0.00002$           & $7490.68697\pm0.00002$            & $7490.68690\pm0.00002$ \\
                        & $P$ [d]               &  $1.33823145\pm0.00000001$        & $1.33823129\pm0.00000001$         & $1.33823126\pm0.00000001$\\\vspace{0.3cm}
                        & $\Delta$BIC                   &  0                           & 0                            & 0 \\

        Orb. Decay   & $T_0$                    & $7490.68721\pm0.00002$            & $7490.68727\pm0.00002$            & $7490.68695\pm0.00002$ \\
                        & $P$ [d]               &  $1.33823132\pm0.00000001$        & $1.33823112\pm0.00000001$         & $1.33823124\pm0.00000001$ \\
                        & $dP/dN$ [d/orbit]     & $(-2.97\pm0.22)\times10^{-10}$    & $(-3.94\pm0.22)\times10^{-10}$    & $(-0.61\pm0.21)\times10^{-10}$ \\\vspace{0.3cm}
                        & $\Delta$BIC                   &  -181.47                           &  -323.87                           & -2.65 \\
        Aps. Prec.  & $T_0$                     & $7490.68608\pm0.00055$            & $7490.68677\pm0.000096$           & $7490.68684\pm0.00018$\\
                            & $P$ [d]           &   $1.33823113\pm0.00000026$       & $1.33823141\pm0.00000008$         & $1.33823142\pm0.00000013$\\
                    & $d\omega/dN$ [rad/orbit]  & $(5.51\pm1.69)\times10^{-4}$      & $(9.69\pm1.16)\times10^{-4}$      & $(6.31\pm2.76)\times10^{-4}$\\
                            & e                 &  $0.0029\pm0.0019$                & $0.0016\pm0.0003$                 & $0.0010\pm0.0007$\\
                            & $\omega_0$ [rad]  &  $2.795\pm0.274$                  & $3.922\pm0.105$                   & $4.455\pm0.515$\\
                            & $\Delta$BIC               &  -177.46                           & -343.83                            & +25.66 \\

         \hline
    \end{tabular}
    \end{adjustwidth}
    \label{tab:MCMC_LTE}
\end{table*}

\begin{figure}
    \centering
    \includegraphics[width=\linewidth]{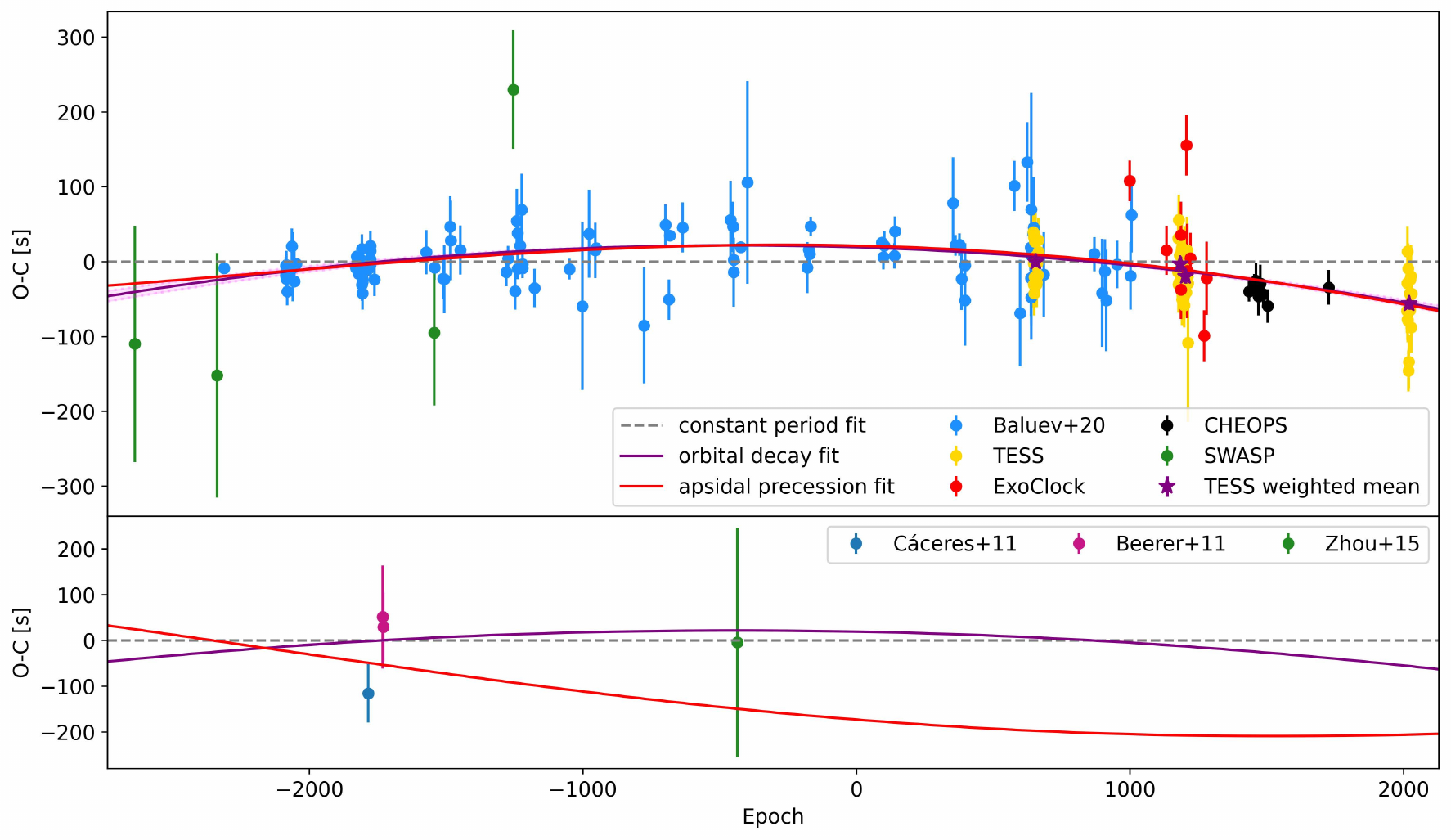}
    \caption{Same as Fig.~\ref{fig:TTVs}, but with the nominal LTE correction applied to the timing data.}
    \label{fig:TTVs_LTE}
\end{figure}

The results in Table~\ref{tab:MCMC_LTE} show that the orbital decay model provides the best fit to the data in case (1), with the nominal LTE correction, although this model is only slightly preferred over the apsidal precession model. In this case, the decay rate would be $\dot{P} = (-7.04\pm0.52)\,$ms\,yr$^{-1}$, leading to $Q_\star'=(4.98\pm0.37)\times10^4$ and a decay time-scale of $\tau = (16.5\pm1.2)\,$Myr. 
The apsidal precession model yields $\dot{\omega} = (0.024\pm0.007)^\circ\,$d$^{-1}$ with an eccentricity of $e = 0.0029\pm0.0019$. This case is displayed in Fig.~\ref{fig:TTVs_LTE}.

In case (2), where the lower boundary of the $1\sigma$ interval of $T_{0,\mathrm{c}}$ is used for the LTE correction, the apsidal precession model is preferred with a $\Delta$BIC of 20 in comparison to the orbital decay model. The linear model is heavily disfavoured. Assuming the orbital decay model to be true, we find an enhanced decay rate of $\dot{P} = (-9.35\pm0.52)$\,ms\,yr$^{-1}$, yielding $Q_\star' = (3.75\pm0.21)\times10^4$ and $\tau = (12.4\pm0.7)$\,Myr.
From the apsidal precession model, we get $\dot{\omega} = (0.041\pm0.005)^\circ\,$d$^{-1}$ with $e = 0.0016\pm0.0003$.
This case is shown in Appendix~\ref{sec:appendixA}, Fig.~\ref{fig:TTVs_LTE_min}.

Case (3), using the upper boundary of the time of inferior conjunction of planet c, shows that the orbital decay model provides the best fit to the data, with the circular orbit model fitting only slightly worse ($\Delta\mathrm{BIC} < 3$), whereas the apsidal precession model fits worse with $\Delta\mathrm{BIC}\approx 25-30$. In this case, the TTV signature is flattened due to the LTE correction. This results in a small decay rate of $\dot{P} = (-1.45\pm0.50)$\,ms\,yr$^{-1}$, leading to $Q_\star' = (2.42\pm0.83)\times10^5$ and $\tau = 80.2\pm27.7$\,Myr.
In the apsidal precession case, we get $\dot{\omega} = (0.027\pm0.012)^\circ\,$d$^{-1}$ with a small eccentricity of $e = 0.0010\pm0.0007$.
The TTV models for this case are shown in Appendix~\ref{sec:appendixA}, Fig.~\ref{fig:TTVs_LTE_max}.

According to \citet{2022AJ....163..281T}, the TTVs induced by planet c onto planet b should not exceed 2\,s, which is why they are neglected here.


\section{Discussion}\label{sec:disc}

\subsection{TTVs of WASP-4\,b}

As is evident from the results of our TTV modelling in Tables~\ref{tab:MCMC} and \ref{tab:MCMC_LTE}, the consideration of the LTE is very important in this system. The results of the orbital decay and apsidal precession modelling are only comparable to each other to a certain degree for the cases that were examined here. 

If the detection of WASP-4\,c should turn out spurious, then the measured decay rate value is comparable, albeit slightly smaller than those from previous studies \citep{2019AJ....157..217B, 2019MNRAS.490.1294B, 2019MNRAS.490.4230S, 2020MNRAS.496L..11B, 2022ApJS..259...62I, 2022AJ....163..281T, 2023A&A...669A.124H}.
In this case, the apsidal precession model would be disfavoured by $\Delta\mathrm{BIC} = 8.5$, which is a bit higher than in the latest study \citep{2023A&A...669A.124H}. The decay rates are consistent within $1\sigma$, with the eccentricity in the precession model being doubled, compared to the previous result, which is still in agreement with our value and uncertainty. The linear model is disfavoured by $\Delta\mathrm{BIC}\approx 110-120$.

The three different LTE corrections lead to three vastly different results. The nominal LTE correction agrees within $3\sigma$ with the results without the LTE correction for most of the parameters. The other solutions do not. Depending on the correction, we get decay rates in the range from $(-9.35\pm0.52)\times10^4$\,ms\,yr$^{-1}$ to $(-1.45\pm0.50)\times10^4$\,ms\,yr$^{-1}$. 
This means that the LTE induced by a planetary companion can solely explain the observed TTVs.
The quite broad range of results can only be further constrained with more radial velocity measurements to reduce the size of the error bars on the parameters of WASP-4\,c. However, a value near the middle of this range is more probable.
From these results, we get modified stellar quality factors in the range from $3.8\times10^4$ to $2.4\times10^5$, which includes the value of $Q_\star' = 1.8\times10^5$ obtained for WASP-12\,b \citep{2020ApJ...888L...5Y}. Moreover, this leads to decay timescales ranging from $12\,$Myr to $80\,$Myr.
The apsidal precession models show a similar range of possible values like the orbital decay models. 
However, with more observations, either transits or especially occultations, the models should be relatively easy to distinguish. Occultation observations with e.g. JWST could perform a double duty in this case. They can help to refine atmospheric properties and provide very accurate occultation timing measurements, which might be able to rule out either of the TTV models.

These results highlight the importance of continued radial velocity observations, even, or especially, in hot Jupiter systems.
One theory of how hot Jupiters get into their tight orbits is high-eccentricity migration, which necessitates the presence of a massive second companion for the excitation of the high eccentricities that are required for this migration pathway \citep{2011Natur.473..187N, 2018ARA&A..56..175D}. Finding such a body requires long observational baselines due to the distance of the perturber to the host star.

\subsection{Optimal observing strategy}
For hot Jupiter systems in general, the most favourable tidal orbital decay systems have, according to Eq.~\ref{eq:Pdot_Q_relation} and Table~2 in \citet{2023A&A...669A.124H}, first and foremost with the highest impact, large stellar radii and short orbital separations. Small stellar tidal quality factors, high planetary masses and low stellar masses are also beneficial. However, stellar tidal quality factors can only be constrained after the observations. Furthermore, orbital periods of the companion bodies smaller than the stellar rotation periods are essential, but this only applies to equatorial orbits.
For polar orbits, orbital decay should always be present, even for fast-rotating stars. Still, slow stellar rotation rates are beneficial here as well.
In summary, as is known, hot Jupiters are prime targets to examine the effect of tidal orbital decay. Moreover, brown dwarfs on close-in orbits should even experience a more pronounced effect due to their higher masses. In addition, companions orbiting evolved or (sub-) giant stars should present the best laboratories to explore orbital decay. Yet, the expected lifetimes of these close-by companions will be relatively short, and observing them is only possible in just the right time window, providing us with only a very small sample of targets (see e.g. \citealt{2022AJ....163..120G}).

In terms of radial velocity monitoring, after sufficient data has been accumulated for the characterisation of hot Jupiters, it would be optimal for the target to be re-observed every few months so that even farther out companions could be detected in principle. This would not only shed light onto the architecture of these systems, but also on their formation and migration pathways.
This case study is an excellent example of this, where an outer companion candidate has been detected \citep{2022AJ....163..281T}, but the whole detection hinges on the latest set of observations to not be spurious or have an incorrect offset in relation to the previously accumulated RV data.
Due to the possibility for the outer planet to induce a small eccentricity onto the orbit of the inner planet, a non-detection constraining possible planet masses and distances from the host star could rule out the apsidal precession models in some cases. This could be achieved by comparing the eccentricity damping timescale to the orbital decay timescale.

For photometric observations, it would be ideal to obtain high-precision transit observations every few months. Long-term monitoring is essential to detect orbital decay signatures with a baseline of more than 15 years having been necessary for WASP-12\,b \citep{2020ApJ...888L...5Y}.
Occultation observations would be even more helpful to differentiate the orbital decay and apsidal precession models, not only from the point-of-view of TTV fitting, but also because they can indicate eccentricities directly. In addition to the timing, secondary eclipse observations can give clues about e.g. the atmopsheric composition or the thermal structure of the atmosphere and the presence of clouds (see e.g. \citealt{2019AJ....157..239J, 2022A&A...668A..17S,2023MNRAS.522.2145V, 2023MNRAS.522.1491S, 2023A&A...675A..81H}).

The PLATO mission \citep{2014ExA....38..249R} has the potential to revolutionise the field with its long, uninterrupted observations. Due to the expected excellent precision of the observations, occultations of hot Jupiters are likely to be detected as well. This should allow discrimination between the orbital decay and apsidal precession models. Besides that, even relatively far-out companions could be detected with radial velocity follow-up from PLATOs ground segment, should they not transit the host star. As well as PLATO, ESA's Gaia mission \citep{2016A&A...595A...1G}, starting with its fourth data release, has the potential to improve our understanding of already known hot Jupiter systems. Its precise astrometric measurements will allow the detection of giant companions in these systems and provide detailed characterisation of their orbits. This will provide information on the formation and migration history of each system.


\section{Conclusions\label{sec:conclusion}}

By analysing new TESS observations of WASP-4 in combination with archival timing data, and taking into account the additional planet candidate in the system, we get a broad range of results for the cause of the TTVs of WASP-4\,b. We examined a total of four cases. The first case does not take into account the presence of planet c, and yields comparable results to the hitherto published results for this system in terms of the orbital decay parameters. In addition, the remaining cases consider, for the first time, the light-time effect from the host star's orbital motion around the system's center of mass, induced by planet c. Depending on the application of the timing correction, we find results that range from a slight preference ($\Delta\mathrm{BIC}\approx4$) of the orbital decay model in the nominal case (using the nominal value for the time of inferior conjuction of planet c from \citet{2022AJ....163..281T}), over a strong preference ($\Delta\mathrm{BIC}\approx20$) of the apsidal precession model using the lower boundary of the respective $1\sigma$ confidence interval, to a nearly indistinguishable result between the linear ephemeris and orbital decay models ($\Delta\mathrm{BIC}\approx2$) using the upper boundary of the respective interval for the time of inferior conjunction of planet c.
These results leave us with no conclusive answer to the question of what the true origin of the TTVs of WASP-4\,b is. 
We need more radial velocity observations to better constrain the phase of planet c. Only then will we be able to determine whether the LTE solely explains the observed TTVs, or whether other mechanisms, like tidal decay or apsidal precession are present.
It is, however, likely to be a mix of all of the effects mentioned here.

This case study highlights the importance of continued monitoring of hot Jupiter systems, in terms of photometric and radial velocity measurements. Only in this way, small effects, like orbital decay or apsidal precession, can be measured and differentiated. An additional benefit is given by the chance of discovering and characterizing companions to hot Jupiters, providing hints of the formation and migration scenarios that lead to these special systems. This is a necessary step to inform our theoretical models of planet formation and migration, and towards the final answer to the question of the origin of these planets.


\vspace{6pt}

\funding{This research was funded by the DFG priority programme SPP 1992 “Exploring the Diversity of Extrasolar Planets (SM 486/2-1)”.}

\dataavailability{The data underlying this article is available at in Appendix~\ref{sec:appendixB}.}



\acknowledgments{This paper includes data collected with the TESS mission, obtained from the MAST data archive at the Space Telescope Science Institute (STScI). Funding for the TESS mission is provided by the NASA Explorer Program. STScI is operated by the Association of Universities for Research in Astronomy, Inc., under NASA contract NAS 5–26555. 
This work made use of Astropy:\footnote{http://www.astropy.org} a community-developed core Python package and an ecosystem of tools and resources for astronomy \citep{2013A&A...558A..33A, 2018AJ....156..123A, 2022ApJ...935..167A}.}

\conflictsofinterest{The authors declare no conflict of interest.}

\appendixtitles{yes} 
\appendixstart
\appendix
\section[\appendixname~\thesection]{TTV fits with LTE correction}\label{sec:appendixA}

\begin{figure}[h]
    \centering
    \includegraphics[width=\linewidth]{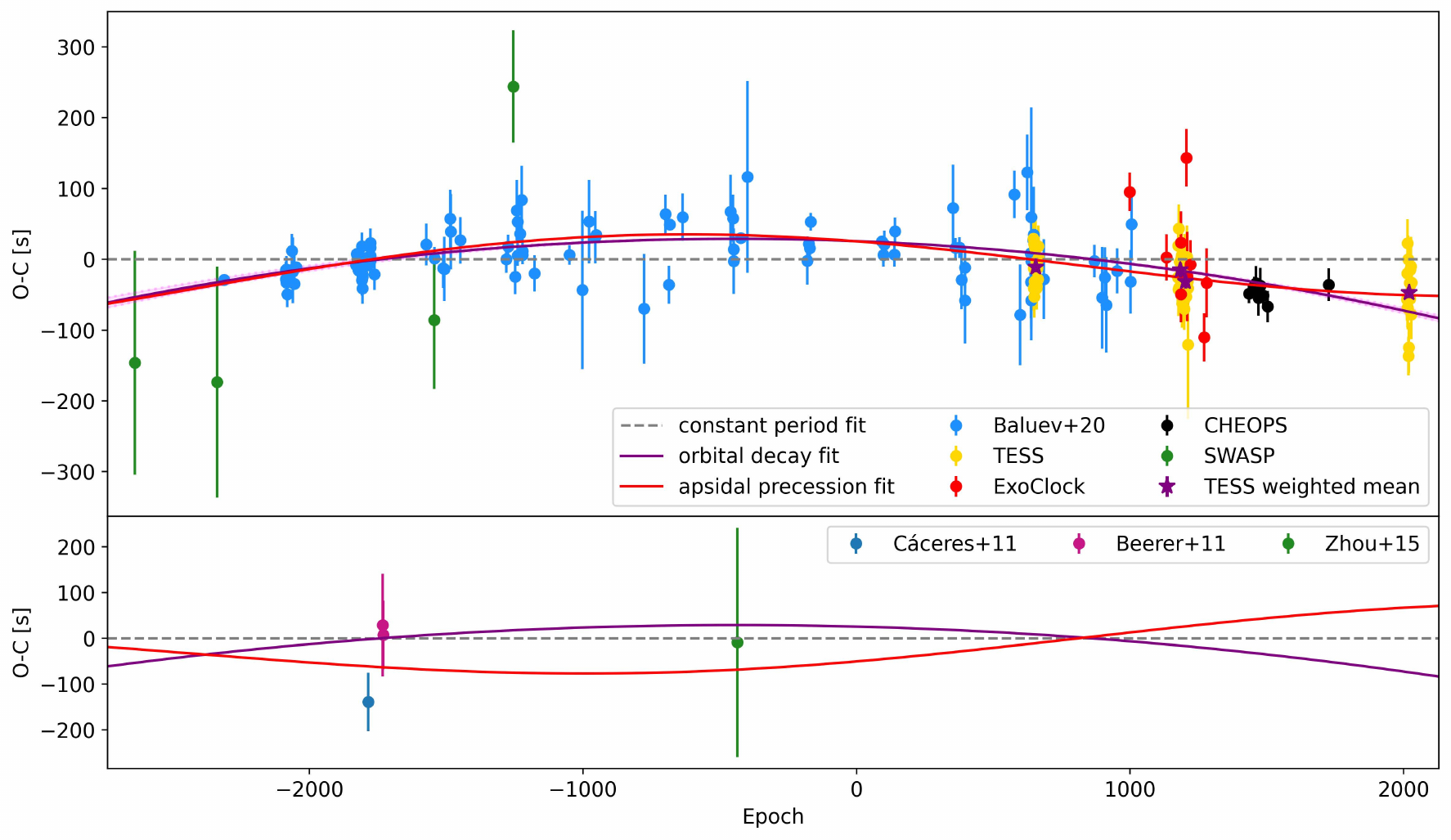}
    \caption{Same as Fig.~\ref{fig:TTVs}, but with the LTE correction applied to the timing data, using the lower limit of the $1\sigma$ interval of the time of inferior conjunction of planet c.}
    \label{fig:TTVs_LTE_min}
\end{figure}

\begin{figure}[h]
    \centering
    \includegraphics[width=\linewidth]{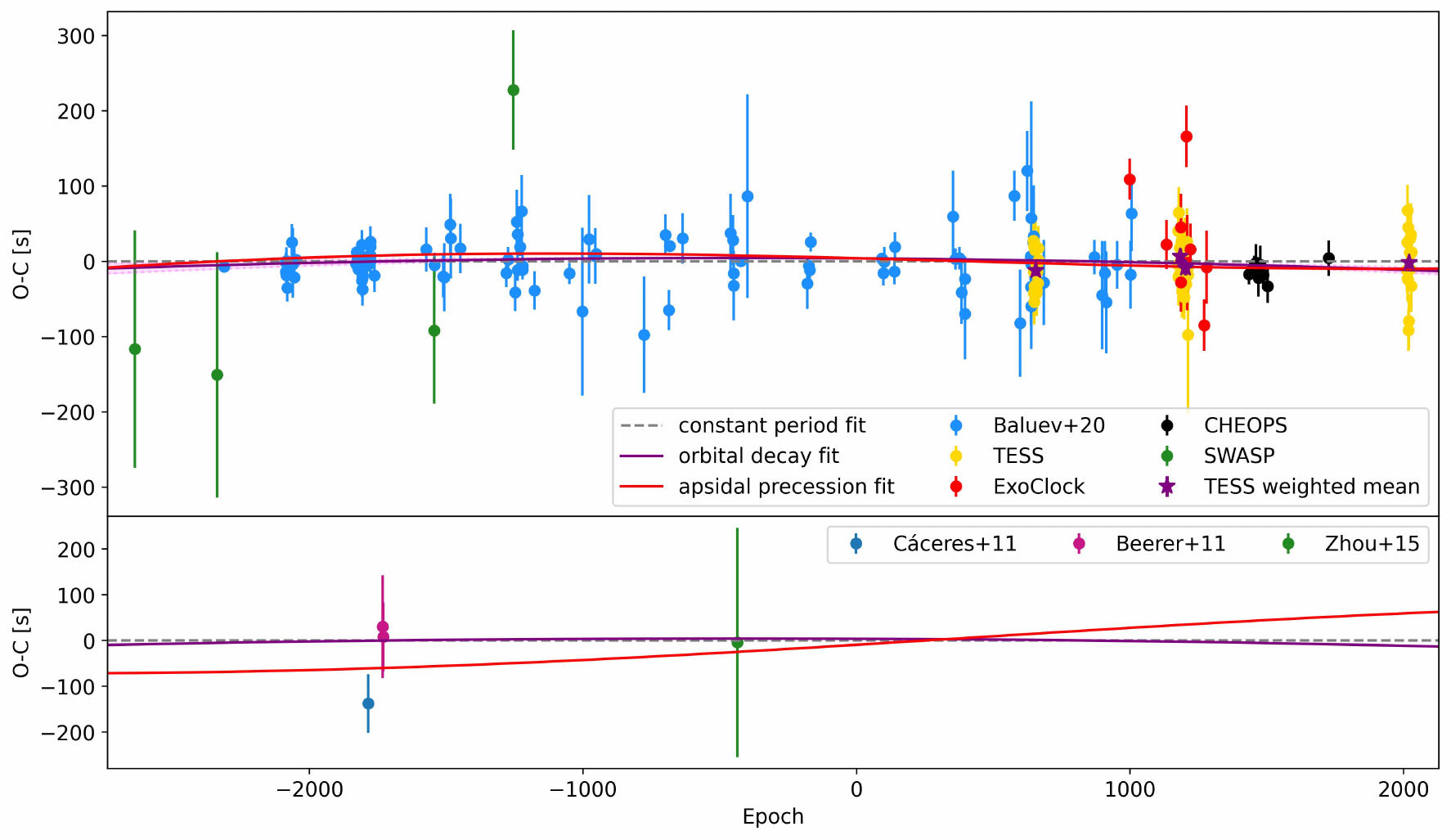}
    \caption{Same as Fig.~\ref{fig:TTVs}, but with the LTE correction, using the upper limit of the $1\sigma$ interval around the nominal time of inferior conjunction of planet c, applied to the timing data.}
    \label{fig:TTVs_LTE_max}
\end{figure}

\section[\appendixname~\thesection]{Transit and Occultation timings}\label{sec:appendixB}

\begin{table}
    \centering
    \setlength{\tabcolsep}{13pt}
    \caption{Transit timings of WASP-4\,b including three different corrections for the LTE. Time is given in BJD$_\mathrm{TDB} - 2450000$. ``Time'' denotes the mid-transit times without any light-time correction, ``Time LTE'' denotes LTE corrected timings using the nominal value for the time of inferior conjunction of planet c from \citet{2022AJ....163..281T}, ``Time LTE$_\mathrm{lower}$'' denotes the timings corrected using the lower boundary of the respective $1\sigma$ interval, and ``LTE$_\mathrm{upper}$'' denotes the timings corrected using upper boundary of the interval. The ``Source'' column denotes the source of the timings, with 0-4 defined as the homogeneous reanalysis from \citet{2020MNRAS.496L..11B} (0), TESS timings (1) from \cite{2023A&A...669A.124H} and this work (last sector), the ExoClock project (2) \citep{2022ApJS..258...40K} as fitted by \citet{2023A&A...669A.124H}, as well as their CHEOPS timings (3), and WASP timings (4), respectively.
    \label{tab:timings}}
    \begin{adjustwidth}{-\extralength}{0cm}
    \begin{tabular}{ c c c c c c c }
    \hline
    Time & Time LTE & Time LTE$_\mathrm{lower}$ & Time LTE$_\mathrm{upper}$ & Error [d] & Epoch & Source \\ \hline
    3960.431480 & 3960.431143 & 3960.431479 & 3960.431127 & 0.001827 & -305  & 4 \\
    4361.900484 & 4361.900090 & 4361.900463 & 4361.900200 & 0.001889 & -5    & 4 \\
    4396.696163 & 4396.695765 & 4396.696138 & 4396.695885 & 0.000085 & 21    & 0 \\
    4697.798154 & 4697.797730 & 4697.798095 & 4697.797934 & 0.000063 & 246   & 0 \\
    4697.798306 & 4697.797882 & 4697.798247 & 4697.798086 & 0.000133 & 246   & 0 \\
    4701.813010 & 4701.812586 & 4701.812951 & 4701.812791 & 0.000217 & 249   & 0 \\
    4701.812980 & 4701.812556 & 4701.812921 & 4701.812761 & 0.000206 & 249   & 0 \\
    4701.812795 & 4701.812371 & 4701.812736 & 4701.812576 & 0.000148 & 249   & 0 \\
    4705.827294 & 4705.826869 & 4705.827234 & 4705.827075 & 0.000210 & 252   & 0 \\
    4728.577932 & 4728.577506 & 4728.577869 & 4728.577718 & 0.000274 & 269   & 0 \\
    4732.592308 & 4732.591883 & 4732.592245 & 4732.592095 & 0.000536 & 272   & 0 \\
    4732.592291 & 4732.591865 & 4732.592227 & 4732.592078 & 0.000145 & 272   & 0 \\
    4740.621727 & 4740.621301 & 4740.621663 & 4740.621516 & 0.000264 & 278   & 0 \\
    4740.621472 & 4740.621046 & 4740.621407 & 4740.621260 & 0.000093 & 278   & 0 \\
    4748.651131 & 4748.650704 & 4748.651065 & 4748.650921 & 0.000056 & 284   & 0 \\
    5041.723762 & 5041.723326 & 5041.723650 & 5041.723608 & 0.000121 & 503   & 0 \\
    5045.738636 & 5045.738201 & 5045.738523 & 5045.738483 & 0.000041 & 506   & 0 \\
    5049.753225 & 5049.752790 & 5049.753112 & 5049.753073 & 0.000042 & 509   & 0 \\
    5053.767759 & 5053.767324 & 5053.767646 & 5053.767608 & 0.000078 & 512   & 0 \\
    5069.826919 & 5069.826484 & 5069.826803 & 5069.826771 & 0.000226 & 524   & 0 \\
    5069.826367 & 5069.825932 & 5069.826251 & 5069.826219 & 0.000249 & 524   & 0 \\
    5069.826591 & 5069.826156 & 5069.826475 & 5069.826443 & 0.000227 & 524   & 0 \\
    5069.826740 & 5069.826305 & 5069.826624 & 5069.826592 & 0.000221 & 524   & 0 \\
    5069.826491 & 5069.826056 & 5069.826374 & 5069.826342 & 0.000128 & 524   & 0 \\
    5073.840928 & 5073.840493 & 5073.840811 & 5073.840781 & 0.000251 & 527   & 0 \\
    5073.841150 & 5073.840715 & 5073.841033 & 5073.841003 & 0.000232 & 527   & 0 \\
    5073.841110 & 5073.840675 & 5073.840993 & 5073.840963 & 0.000168 & 527   & 0 \\
    5073.841123 & 5073.840688 & 5073.841006 & 5073.840976 & 0.000209 & 527   & 0 \\
    5096.591358 & 5096.590923 & 5096.591237 & 5096.591215 & 0.000148 & 544   & 0 \\
    5100.605932 & 5100.605497 & 5100.605810 & 5100.605789 & 0.000101 & 547   & 0 \\
    5112.650371 & 5112.649936 & 5112.650247 & 5112.650230 & 0.000236 & 556   & 0 \\
    5112.650296 & 5112.649861 & 5112.650172 & 5112.650156 & 0.000177 & 556   & 0 \\
    5112.650162 & 5112.649727 & 5112.650038 & 5112.650021 & 0.000207 & 556   & 0 \\
    5112.650086 & 5112.649651 & 5112.649962 & 5112.649945 & 0.000261 & 556   & 0 \\
    5132.723325 & 5132.722891 & 5132.723198 & 5132.723188 & 0.000257 & 571   & 0 \\
    5385.649482 & 5385.649056 & 5385.649307 & 5385.649388 & 0.000342 & 760   & 0 \\
    5424.456950 & 5424.456526 & 5424.456768 & 5424.456862 & 0.001127 & 789   & 4 \\
    5425.796186 & 5425.795762 & 5425.796004 & 5425.796099 & 0.000188 & 790   & 0 \\
    5468.619420 & 5468.618999 & 5468.619229 & 5468.619339 & 0.000320 & 822   & 0 \\
    5473.972334 & 5473.971913 & 5473.972142 & 5473.972254 & 0.000524 & 826   & 0 \\
    5502.076007 & 5502.075589 & 5502.075810 & 5502.075932 & 0.000471 & 847   & 0 \\
    5506.090489 & 5506.090071 & 5506.090291 & 5506.090414 & 0.000616 & 850   & 0 \\
    5551.590205 & 5551.589791 & 5551.589999 & 5551.590137 & 0.000378 & 884   & 0 \\
    5777.750955 & 5777.750564 & 5777.750705 & 5777.750916 & 0.000220 & 1053  & 0 \\
    5785.780551 & 5785.780160 & 5785.780299 & 5785.780512 & 0.000196 & 1059  & 0 \\
    5811.209560 & 5811.209172 & 5811.209303 & 5811.209524 & 0.000918 & 1078  & 4 \\
    5820.574064 & 5820.573678 & 5820.573805 & 5820.574029 & 0.000286 & 1085  & 0 \\
    5828.604540 & 5828.604155 & 5828.604280 & 5828.604506 & 0.000493 & 1091  & 0 \\
    \hline
    \end{tabular}
    \end{adjustwidth}
\end{table}

\setcounter{table}{0}
\begin{table}
    \centering
    \setlength{\tabcolsep}{13pt}
    \caption{\textit{continued.}}
    \begin{adjustwidth}{-\extralength}{0cm}
    \begin{tabular}{ c c c c c c c }
    \hline
    Time & Time LTE & Time LTE$_\mathrm{lower}$ & Time LTE$_\mathrm{upper}$ & Error & Epoch & Source \\ \hline
    5832.618492 & 5832.618107 & 5832.618231 & 5832.618458 & 0.000163 & 1094  & 0 \\
    5832.619044 & 5832.618660 & 5832.618783 & 5832.619011 & 0.000426 & 1094  & 0 \\
    5844.662930 & 5844.662547 & 5844.662667 & 5844.662898 & 0.000138 & 1103  & 0 \\
    5852.692871 & 5852.692489 & 5852.692606 & 5852.692840 & 0.000557 & 1109  & 0 \\
    5856.706724 & 5856.706342 & 5856.706458 & 5856.706693 & 0.000210 & 1112  & 0 \\
    5856.706666 & 5856.706284 & 5856.706400 & 5856.706635 & 0.000098 & 1112  & 0 \\
    5915.588533 & 5915.588159 & 5915.588257 & 5915.588508 & 0.000297 & 1156  & 0 \\
    6086.882427 & 6086.882078 & 6086.882118 & 6086.882414 & 0.000162 & 1284  & 0 \\
    6149.778721 & 6149.778382 & 6149.778402 & 6149.778713 & 0.001294 & 1331  & 0 \\
    6181.897390 & 6181.897056 & 6181.897065 & 6181.897383 & 0.000680 & 1355  & 0 \\
    6212.676454 & 6212.676126 & 6212.676124 & 6212.676449 & 0.000428 & 1378  & 0 \\
    6216.691180 & 6216.690852 & 6216.690849 & 6216.691175 & 0.000067 & 1381  & 0 \\
    6450.880446 & 6450.880160 & 6450.880079 & 6450.880446 & 0.000895 & 1556  & 0 \\
    6556.602268 & 6556.602001 & 6556.601887 & 6556.602267 & 0.000315 & 1635  & 0 \\
    6572.659887 & 6572.659624 & 6572.659504 & 6572.659886 & 0.000311 & 1647  & 0 \\
    6576.675568 & 6576.675305 & 6576.675184 & 6576.675567 & 0.000055 & 1650  & 0 \\
    6639.572561 & 6639.572310 & 6639.572170 & 6639.572558 & 0.000388 & 1697  & 0 \\
    6873.763138 & 6873.762933 & 6873.762724 & 6873.763123 & 0.000602 & 1872  & 0 \\
    6885.807112 & 6885.806910 & 6885.806696 & 6885.807096 & 0.000386 & 1881  & 0 \\
    6889.821295 & 6889.821093 & 6889.820879 & 6889.821279 & 0.000198 & 1884  & 0 \\
    6889.821104 & 6889.820902 & 6889.820688 & 6889.821088 & 0.000535 & 1884  & 0 \\
    6924.615500 & 6924.615305 & 6924.615081 & 6924.615481 & 0.000086 & 1910  & 0 \\
    6956.734051 & 6956.733862 & 6956.733630 & 6956.734030 & 0.001568 & 1934  & 0 \\
    7249.805368 & 7249.805234 & 7249.804933 & 7249.805315 & 0.000394 & 2153  & 0 \\
    7257.835029 & 7257.834897 & 7257.834595 & 7257.834976 & 0.000082 & 2159  & 0 \\
    7261.849657 & 7261.849525 & 7261.849222 & 7261.849603 & 0.000129 & 2162  & 0 \\
    7265.864781 & 7265.864650 & 7265.864347 & 7265.864726 & 0.000147 & 2165  & 0 \\
    7613.804649 & 7613.804575 & 7613.804219 & 7613.804542 & 0.000084 & 2425  & 0 \\
    7621.833815 & 7621.833742 & 7621.833385 & 7621.833706 & 0.000191 & 2431  & 0 \\
    7625.848681 & 7625.848609 & 7625.848252 & 7625.848572 & 0.000226 & 2434  & 0 \\
    7675.363083 & 7675.363018 & 7675.362656 & 7675.362966 & 0.000196 & 2471  & 0 \\
    7679.378156 & 7679.378091 & 7679.377730 & 7679.378038 & 0.000228 & 2474  & 0 \\
    7961.745397 & 7961.745366 & 7961.744993 & 7961.745227 & 0.000707 & 2685  & 0 \\
    7973.788824 & 7973.788795 & 7973.788421 & 7973.788652 & 0.000162 & 2694  & 0 \\
    7993.862304 & 7993.862276 & 7993.861903 & 7993.862128 & 0.000172 & 2709  & 0 \\
    8004.567627 & 8004.567600 & 8004.567227 & 8004.567449 & 0.000481 & 2717  & 0 \\
    8020.626070 & 8020.626045 & 8020.625672 & 8020.625890 & 0.000702 & 2729  & 0 \\
    8020.626608 & 8020.626583 & 8020.626210 & 8020.626427 & 0.000345 & 2729  & 0 \\
    8262.847714 & 8262.847706 & 8262.847347 & 8262.847486 & 0.000387 & 2910  & 0 \\
    8290.948607 & 8290.948600 & 8290.948243 & 8290.948373 & 0.000824 & 2931  & 0 \\
    8325.744960 & 8325.744955 & 8325.744601 & 8325.744720 & 0.000617 & 2957  & 0 \\
    8341.802413 & 8341.802409 & 8341.802057 & 8341.802170 & 0.000361 & 2969  & 0 \\
    8343.140176 & 8343.140171 & 8343.139820 & 8343.139932 & 0.000591 & 2970  & 0 \\
    8345.816340 & 8345.816336 & 8345.815985 & 8345.816096 & 0.000657 & 2972  & 0 \\
    8345.817698 & 8345.817694 & 8345.817343 & 8345.817454 & 0.001797 & 2972  & 0 \\
    8345.816987 & 8345.816983 & 8345.816632 & 8345.816743 & 0.000103 & 2972  & 0 \\
    8349.831750 & 8349.831747 & 8349.831396 & 8349.831506 & 0.000212 & 2975  & 0 \\
    8353.846717 & 8353.846713 & 8353.846363 & 8353.846471 & 0.000650 & 2978  & 0 \\
    8357.861497 & 8357.861493 & 8357.861144 & 8357.861251 & 0.000783 & 2981  & 0 \\
    8357.861077 & 8357.861073 & 8357.860723 & 8357.860830 & 0.000105 & 2981  & 0 \\
    8406.037105 & 8406.037103 & 8406.036759 & 8406.036849 & 0.000654 & 3017  & 0 \\
    \hline
    \end{tabular}
    \end{adjustwidth}
\end{table}

\setcounter{table}{0}
\begin{table}
    \centering
    \setlength{\tabcolsep}{13pt}
    \caption{\textit{continued.}}
    \begin{adjustwidth}{-\extralength}{0cm}
    \begin{tabular}{ c c c c c c c }
    \hline
    Time & Time LTE & Time LTE$_\mathrm{lower}$ & Time LTE$_\mathrm{upper}$ & Error & Epoch & Source \\ \hline
    8355.184967 & 8355.184963 & 8355.184613 & 8355.184721 & 0.000330 & 2979  & 1 \\
    8356.522391 & 8356.522388 & 8356.522038 & 8356.522145 & 0.000367 & 2980  & 1 \\
    8357.860950 & 8357.860947 & 8357.860597 & 8357.860704 & 0.000316 & 2981  & 1 \\
    8359.199579 & 8359.199576 & 8359.199226 & 8359.199333 & 0.000311 & 2982  & 1 \\
    8360.536947 & 8360.536944 & 8360.536594 & 8360.536700 & 0.000343 & 2983  & 1 \\
    8361.875428 & 8361.875424 & 8361.875075 & 8361.875180 & 0.000307 & 2984  & 1 \\
    8363.214209 & 8363.214205 & 8363.213856 & 8363.213961 & 0.000361 & 2985  & 1 \\
    8364.551875 & 8364.551871 & 8364.551522 & 8364.551627 & 0.000348 & 2986  & 1 \\
    8365.890728 & 8365.890725 & 8365.890376 & 8365.890480 & 0.000393 & 2987  & 1 \\
    8369.905146 & 8369.905143 & 8369.904795 & 8369.904898 & 0.000361 & 2990  & 1 \\
    8371.242934 & 8371.242931 & 8371.242583 & 8371.242685 & 0.000310 & 2991  & 1 \\
    8372.581201 & 8372.581198 & 8372.580850 & 8372.580952 & 0.000384 & 2992  & 1 \\
    8373.919748 & 8373.919745 & 8373.919397 & 8373.919499 & 0.000348 & 2993  & 1 \\
    8375.257960 & 8375.257957 & 8375.257609 & 8375.257710 & 0.000319 & 2994  & 1 \\
    8376.596324 & 8376.596321 & 8376.595974 & 8376.596074 & 0.000336 & 2995  & 1 \\
    8377.934241 & 8377.934238 & 8377.933891 & 8377.933991 & 0.000352 & 2996  & 1 \\
    8379.273009 & 8379.273006 & 8379.272659 & 8379.272759 & 0.000344 & 2997  & 1 \\
    8380.611058 & 8380.611055 & 8380.610708 & 8380.610807 & 0.000374 & 2998  & 1 \\
    8653.610237 & 8653.610237 & 8653.609933 & 8653.609935 & 0.000265 & 3202  & 0 \\
    8692.418351 & 8692.418350 & 8692.418054 & 8692.418042 & 0.000833 & 3231  & 0 \\
    8705.801002 & 8705.801000 & 8705.800708 & 8705.800691 & 0.000492 & 3241  & 0 \\
    8712.491707 & 8712.491705 & 8712.491413 & 8712.491394 & 0.000782 & 3246  & 0 \\
    8764.683292 & 8764.683288 & 8764.683008 & 8764.682971 & 0.000366 & 3285  & 0 \\
    8827.581466 & 8827.581460 & 8827.581194 & 8827.581134 & 0.000314 & 3332  & 2 \\
    8831.594691 & 8831.594684 & 8831.594420 & 8831.594358 & 0.000521 & 3335  & 0 \\
    8835.610326 & 8835.610320 & 8835.610056 & 8835.609993 & 0.000581 & 3338  & 0 \\
    9006.903416 & 9006.903399 & 9006.903179 & 9006.903056 & 0.000379 & 3466  & 2 \\
    9063.108812 & 9063.108790 & 9063.108585 & 9063.108444 & 0.000397 & 3508  & 1 \\
    9064.447561 & 9064.447539 & 9064.447335 & 9064.447192 & 0.000362 & 3509  & 1 \\
    9065.785080 & 9065.785058 & 9065.784854 & 9065.784711 & 0.000439 & 3510  & 1 \\
    9067.124306 & 9067.124284 & 9067.124080 & 9067.123937 & 0.000390 & 3511  & 1 \\
    9068.461570 & 9068.461548 & 9068.461345 & 9068.461201 & 0.000417 & 3512  & 1 \\
    9069.799976 & 9069.799954 & 9069.799751 & 9069.799607 & 0.000463 & 3513  & 1 \\
    9071.138455 & 9071.138432 & 9071.138230 & 9071.138085 & 0.000530 & 3514  & 1 \\
    9076.491113 & 9076.491090 & 9076.490889 & 9076.490743 & 0.000431 & 3518  & 1 \\
    9077.829080 & 9077.829057 & 9077.828856 & 9077.828709 & 0.000453 & 3519  & 2 \\
    9077.829928 & 9077.829905 & 9077.829704 & 9077.829557 & 0.000511 & 3519  & 2 \\
    9077.829278 & 9077.829255 & 9077.829054 & 9077.828908 & 0.000418 & 3519  & 1 \\
    9079.167681 & 9079.167658 & 9079.167458 & 9079.167310 & 0.000403 & 3520  & 1 \\
    9080.506153 & 9080.506130 & 9080.505930 & 9080.505782 & 0.000467 & 3521  & 1 \\
    9081.844421 & 9081.844397 & 9081.844198 & 9081.844050 & 0.000441 & 3522  & 1 \\
    9083.181857 & 9083.181834 & 9083.181635 & 9083.181486 & 0.000398 & 3523  & 1 \\
    9084.520535 & 9084.520512 & 9084.520313 & 9084.520164 & 0.000599 & 3524  & 1 \\
    9088.534731 & 9088.534707 & 9088.534509 & 9088.534359 & 0.000346 & 3527  & 1 \\
    9089.873673 & 9089.873649 & 9089.873452 & 9089.873301 & 0.000309 & 3528  & 1 \\
    9091.211733 & 9091.211709 & 9091.211512 & 9091.211361 & 0.000347 & 3529  & 1 \\
    9092.549788 & 9092.549764 & 9092.549568 & 9092.549416 & 0.000379 & 3530  & 1 \\
    9093.887619 & 9093.887595 & 9093.887399 & 9093.887247 & 0.000341 & 3531  & 1 \\
    9095.226017 & 9095.225992 & 9095.225796 & 9095.225644 & 0.000364 & 3532  & 1 \\
    9096.564853 & 9096.564829 & 9096.564633 & 9096.564481 & 0.000453 & 3533  & 1 \\
    9097.902814 & 9097.902789 & 9097.902594 & 9097.902441 & 0.000368 & 3534  & 1 \\
    \hline
    \end{tabular}
    \end{adjustwidth}
\end{table}

\setcounter{table}{0}
\begin{table}
    \centering
    \setlength{\tabcolsep}{13pt}
    \caption{\textit{continued.}}
    \begin{adjustwidth}{-\extralength}{0cm}
    \begin{tabular}{ c c c c c c c }
    \hline
    Time & Time LTE & Time LTE$_\mathrm{lower}$ & Time LTE$_\mathrm{upper}$ & Error & Epoch & Source \\ \hline
    9103.255438 & 9103.255413 & 9103.255219 & 9103.255064 & 0.000366 & 3538  & 1 \\
    9104.594195 & 9104.594169 & 9104.593976 & 9104.593821 & 0.000334 & 3539  & 1 \\
    9104.595944 & 9104.595919 & 9104.595726 & 9104.595570 & 0.000471 & 3539  & 2 \\
    9105.932501 & 9105.932475 & 9105.932283 & 9105.932127 & 0.000440 & 3540  & 1 \\
    9107.270785 & 9107.270759 & 9107.270567 & 9107.270410 & 0.000515 & 3541  & 1 \\
    9108.608905 & 9108.608879 & 9108.608687 & 9108.608530 & 0.000499 & 3542  & 1 \\
    9108.608695 & 9108.608669 & 9108.608477 & 9108.608321 & 0.000731 & 3542  & 2 \\
    9109.946842 & 9109.946816 & 9109.946625 & 9109.946467 & 0.000361 & 3543  & 1 \\
    9111.284969 & 9111.284943 & 9111.284752 & 9111.284594 & 0.000350 & 3544  & 1 \\
    9112.622279 & 9112.622253 & 9112.622062 & 9112.621904 & 0.001219 & 3545  & 1 \\
    9124.667672 & 9124.667644 & 9124.667457 & 9124.667295 & 0.000395 & 3554  & 2 \\
    9191.578053 & 9191.578019 & 9191.577852 & 9191.577668 & 0.000395 & 3604  & 2 \\
    9203.623026 & 9203.622991 & 9203.622827 & 9203.622639 & 0.000563 & 3613  & 2 \\
    9411.048723 & 9411.048663 & 9411.048564 & 9411.048314 & 0.000158 & 3768  & 3 \\
    9436.475245 & 9436.475181 & 9436.475090 & 9436.474834 & 0.000150 & 3787  & 3 \\
    9444.504679 & 9444.504614 & 9444.504526 & 9444.504267 & 0.000282 & 3793  & 3 \\
    9457.886752 & 9457.886685 & 9457.886602 & 9457.886339 & 0.000291 & 3803  & 3 \\
    9465.916342 & 9465.916274 & 9465.916193 & 9465.915928 & 0.000290 & 3809  & 3 \\
    9480.636724 & 9480.636653 & 9480.636577 & 9480.636308 & 0.000219 & 3820  & 3 \\
    9502.048249 & 9502.048175 & 9502.048106 & 9502.047832 & 0.000257 & 3836  & 3 \\
    9800.474200 & 9800.474078 & 9800.474109 & 9800.473767 & 0.000270 & 4059  & 3 \\
    10183.208105 & 10183.207912 & 10183.208066 & 10183.207674 & 0.000341 & 4345  & 1 \\
    10184.546761 & 10184.546568 & 10184.546722 & 10184.546330 & 0.000307 & 4346  & 1 \\
    10185.884435 & 10185.884242 & 10185.884396 & 10185.884004 & 0.000325 & 4347  & 1 \\
    10187.223718 & 10187.223524 & 10187.223680 & 10187.223287 & 0.000391 & 4348  & 1 \\
    10188.560885 & 10188.560691 & 10188.560847 & 10188.560454 & 0.000344 & 4349  & 1 \\
    10189.899918 & 10189.899724 & 10189.899880 & 10189.899488 & 0.000386 & 4350  & 1 \\
    10191.236568 & 10191.236374 & 10191.236530 & 10191.236138 & 0.000318 & 4351  & 1 \\
    10192.574939 & 10192.574744 & 10192.574901 & 10192.574509 & 0.000410 & 4352  & 1 \\
    10196.590687 & 10196.590492 & 10196.590650 & 10196.590257 & 0.000295 & 4355  & 1 \\
    10197.929131 & 10197.928935 & 10197.929094 & 10197.928701 & 0.000376 & 4356  & 1 \\
    10199.266892 & 10199.266696 & 10199.266855 & 10199.266462 & 0.000348 & 4357  & 1 \\
    10200.605601 & 10200.605405 & 10200.605564 & 10200.605171 & 0.000352 & 4358  & 1 \\
    10201.943596 & 10201.943399 & 10201.943559 & 10201.943166 & 0.000314 & 4359  & 1 \\
    10203.282118 & 10203.281921 & 10203.282081 & 10203.281688 & 0.000479 & 4360  & 1 \\
    10204.619557 & 10204.619360 & 10204.619521 & 10204.619127 & 0.000398 & 4361  & 1 \\
    10205.958311 & 10205.958114 & 10205.958275 & 10205.957881 & 0.000370 & 4362  & 1 \\
    \hline
    \end{tabular}
    \end{adjustwidth}
\end{table}

\begin{table}[H]
    \centering
    \setlength{\tabcolsep}{13pt}
    \caption{Occultation timings of WASP-4\,b including three different corrections for the LTE. Time is given in BJD$_\mathrm{TDB} - 2450000$. ``Time'' denotes the mid-occultation times without the light-time correction due to planet c, ``Time LTE'' denotes LTE corrected timings using the nominal value for the time of inferior conjunction of planet c from \citet{2022AJ....163..281T}, ``Time LTE$_\mathrm{lower}$'' denotes the timings corrected using the lower boundary of the respective $1\sigma$ interval, and ``LTE$_\mathrm{upper}$'' denotes the timings corrected using upper boundary of the interval. The ``Source'' column denotes the source of the timings, with (5) the timing from \citet{2011A&A...530A...5C}, (6) the two from \citet{2011ApJ...727...23B}, and (7) the one from \citet{2015MNRAS.454.3002Z}.
    \label{tab:timings_occ}}
    \begin{adjustwidth}{-\extralength}{0cm}
    \begin{tabular}{ c c c c c c c }
    \hline
    Time & Time LTE & Time LTE$_\mathrm{lower}$ & Time LTE$_\mathrm{upper}$ & Error [d] & Epoch & Source \\ \hline
    5102.611620 & 5102.611185 & 5102.611498 & 5102.611478 & 0.000740 & 548.5 & 5 \\
    5172.201590 & 5172.201156 & 5172.201456 & 5172.201460 & 0.001300 & 600.5 & 6 \\
    5174.877800 & 5174.877366 & 5174.877665 & 5174.877671 & 0.000870 & 602.5 & 6 \\
    6907.887140 & 6907.886942 & 6907.886723 & 6907.887123 & 0.002900 & 1897.5 & 7 \\
    \hline
    \end{tabular}
    \end{adjustwidth}
\end{table}

\begin{adjustwidth}{-\extralength}{0cm}

\reftitle{References}


\bibliography{main}


%


\PublishersNote{}
\end{adjustwidth}
\end{document}